# A NEW DNA BASED APPROACH OF GENERATING KEY-DEPENDENT SHIFTROWS TRANSFORMATION

Auday H. Al-Wattar[1], Ramlan Mahmod[2], Zuriati Ahmad Zukarnain3, and Nur Izura Udzir4,

[1]Faculty of Computer Science and Information Technology
Universiti Putra Malaysia,
43400 UPM, Serdang, Selangor.

ahsa.alwattar@gmail.com    Ramaln@upm.edu.my
Zuraiti@upm.edu.my         Izura@upm.edu.my

## ABSTRACT

*The use of key-dependent shiftRows can be considered as one of the applied methods for altering the quality of a cryptographic algorithm. This article describes one approach for changing the ShiftRows transformation employed in the algorithm AES. The approach employs methods inspired from DNA processes and structure which depended on the key while the parameters of the created new ShiftRows have characteristics identical to those of the original algorithm AES in addition to increase its resistance against attacks. The proposed new ShiftRows were tested for coefficient correlation for dynamic and static independence between the input and output. The NIST Test Suite tests were used to test the randomness for the block cipher that used the new transformation.*

## 1. INTRODUCTION

Through the rapid expansion of the Internet and increasing the reliance upon in all areas of life, the need for a highly effective way to achieve security is fateful and crucial. Cryptography has been and remains the most efficient approach used to achieve security. Rijndael is a symmetric key block cipher that was chosen by (NIST) (National Institute of Standards and Technology), [1] in 2001 as (advanced Encryption Standard, FIPS 197) AES.This encryption is dependent on substitution-permutation network Shannon (SPN).

Generally, it is based on repeated rounds of transformation that convert input plaintext to ciphertext (encrypted text) output. Each round includes several procedures and always involves a relying on the private cryptographic key. Multiple cycles determine inverse transforming ciphertext into the original, using the same cryptographic key. AES has fixed block size (128-bit), and a key length is 128, 192 or 256 bits, relating to the number of rounds for the algorithm. It is runs a byte array called state of 4x4 size in each round of encryption / decryption methods. Most of the algorithm calculations are achieved in finite fields. [2, 3]

In AES, the ShiftRows is transformation is one of linear unit of symmetric encryption algorithms.It is a transposition action where each row of the state cyclically shifted a number of times. The aim of this function is to scramble the byte order within each 128-bit block, to supply diffusion of bits across multiple rounds. This transformation includes shifting the state rows as: The first row in the state does not move, the second row is circular left shifted by 1 byte, the third row is circular left shifted by 2 bytes, and the last row is circular left shifted by 3 bytes. Figure 1 shows the AES ShiftRows transformation.

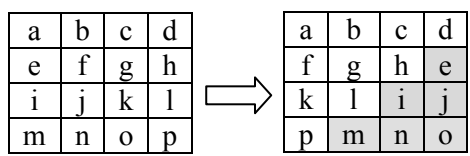

Figure 1. AES ShitRows transformation

According to [4], AES without ShiftRows stage are more than that of AES, their values differ very little for different rounds. This is assign of poor encryption quality

The most challenging for AES algorithm is the linear and differential cryptanalysis where, Rijndael can be issue to standard techniques of differential and linear cryptanalysis.

From the analysis of resistance against differential and linear cryptanalysis, it was found that the arbitrary unknown and key-dependent substitution and permutation transformations is consider as a good factor in enhancing the resistance of block cipher against the differential and linear attacks, since the differential and linear attacks need known transformations. The specific properties of substitution and permutation functions, specifically the structure that is entirely dynamic and unknown to the cryptanalyst, aid and support the block cipher to be resistance against attacks. For the attacker, differential and linear trail over multiple rounds is consider as a vital requirement, and in the existence of key-dependent dynamic transformations output differences rely on additional key used. In addition for any additional different key values, there are unrelated differentials over multiple rounds and therefore diverse linear differential trails. As result it is difficult for the attacker to utilize the current linear and differential techniques of cryptanalysis.

Although all previous works on enhancing the security of AES block cipher against attacks, no one has proposed or suggested a ShiftRows that is designed or created using DNA structure and bio-inspired techniques.

This paper proposed a novel technique for gaining a powerful key-dependent ShiftRows based on operations that have been inspired from really biological DNA structure and processes. Subsequently, it tested the new ShiftRows using the correlation coefficient test, the NIST randomness tests for the cryptosystem that used the DNA-based ShifRows and finally performed a cryptanalysis for the proposed transformation.

## 2. DNA BACKGROUND

DNA (Deoxyribo Nucleic Acid) is a molecule that represents the genetic material for all living organisms. It is the information carrier of all life forms, and considered as the genetic drawing of living or existing creatures. DNA contains two such chains, twisted around each other to form a double-stranded helix with the bases on the inside.

A single-strand of DNA consists of a chain of molecules known as bases, defined as four characters {A, C, G, and T} [5, 6]. One of the most basic attributes of the DNA strand series is that it is oriented; consequently, CGTAGGA is distinct from AGGATGC.

The DNA strands subsist as pairs as (A) associated with (T) and (C) associated with (G) forming units named base pairs, as shown in Figure 2. The reverse DNA strand represents the opposite of the strand bases; for example, GCATAA becomes AATACG, while the complement of the strands can be represented as $\overline{A} \equiv T$ and $\overline{C} \equiv G$. Accordingly, the reverse complementary of the strands can be represented in two actions: reverse the basis of the strands and the second complements the resulting bases, such that (A) becomes (T) and (C) becomes (G) and vice versa. Therefore, GCATAA will become TTATGC [7]. The reverse complement of the DNA strand bases is one of the main methods that characterize the biology DNA system. The reverse complement method is composed of two stages. The first stage is the reverse which means reverse the order of the DNA bases since the DNA strands of DNA sequences which hold the bases are oriented such that they have two opposite sides known as 3' end and 5' end [8] .All the processes that take place over DNA occurs from the 3' end to the 5' end, and verse versa, accordingly, TTCA is distinct from ACTT. The second stage is the complement which is comparable to the complement operation in binary code but within the letters rather than digits Thus the reverse complement of AGCTTGAC is GTCAAGCT [9]. In this paper, the reverse complement process will be used as inspiration for the new key-dependent ShiftRows transformation.

| DNA strand | Reverse | Reverse Complements |
|---|---|---|
| GCATAA | AATACG | TTATGC |

Figure 2. DNA bases reverse complement technique

## 3. THE PROPOSED METHOD

In this paper the proposed ShiftRows transformation will refer to as a key-dependent dynamic transformation ($KdD\_T_r$).

The $KdD\_T_r$ process is inspired by DNA-strands structures including its nature, orientation as well as DNA-bases and reverse-complement process.

In this transformation the cipher key *Kr* at round r is used as key for applying the reverse-complement over state bytes (byte level). This technique makes the byte transposition process dynamic rather than static, depending on the *Kr*. Four specific bytes of *Kr* will be used to obtain four values ranging between 0 and 3, in such a way that each single key value (N1, N4) is obtained by one byte. According to key value the number of bytes that form the DNA-base will be specified. Note that in this transformation the DNA-bases will be represented by a variable number of bytes instead of two bits as previously. The key value (N1, N1) would specify the number of bytes that form one DNA-base. The state STE is considered as four DNA-strands such that each row forms one DNA-strand, but the number of DNA-bases that form each DNA-strand differs depending on key value.

Every key (N1, N4) has four values, where each value represents the number of bytes for single STE row (DNA-strand) that can be used to represent a single DNA base. The cases of key are shown in Table 1,these values of the key will be used to specify the mechanism of byte level

Table 1: Transposition key values

| value | meaning |
|---|---|
| 0 | Each single byte of STE row represent a base |
| 1 | No value/ leave bytes of STE row without any change |
| 2 | Ever two byte s of STE row represent a base |
| 3 | All the 4 STE row bytes represent a single base |

reverse-complement of state rows, where a, b, c and, d are 4 bytes, representing a single row of stat STE; note that a single row represents a single DNA-strand. Every DNA-base can be represented by one, two or four bytes, according to the key values and after the representing is done, the reverse-complement method is performed over the resulting DNA-bases. If every DNA-base is formed by one byte (key value = 0) then the reverse complement will be achieved on the state row in byte by byte level; however if one DNA-base is formed by two bytes (key value = 2) then the reverse complement will be done on state row as every two bytes, Finally, if every DNA-base is formed using four bytes (key value = 3) then the reverse-complement will be achieved on the whole row as one part. There is only one case (key value =1) where there is no reverse but, just complement for the state row. Note that each row has a unique key value as illustrated in Figure 3

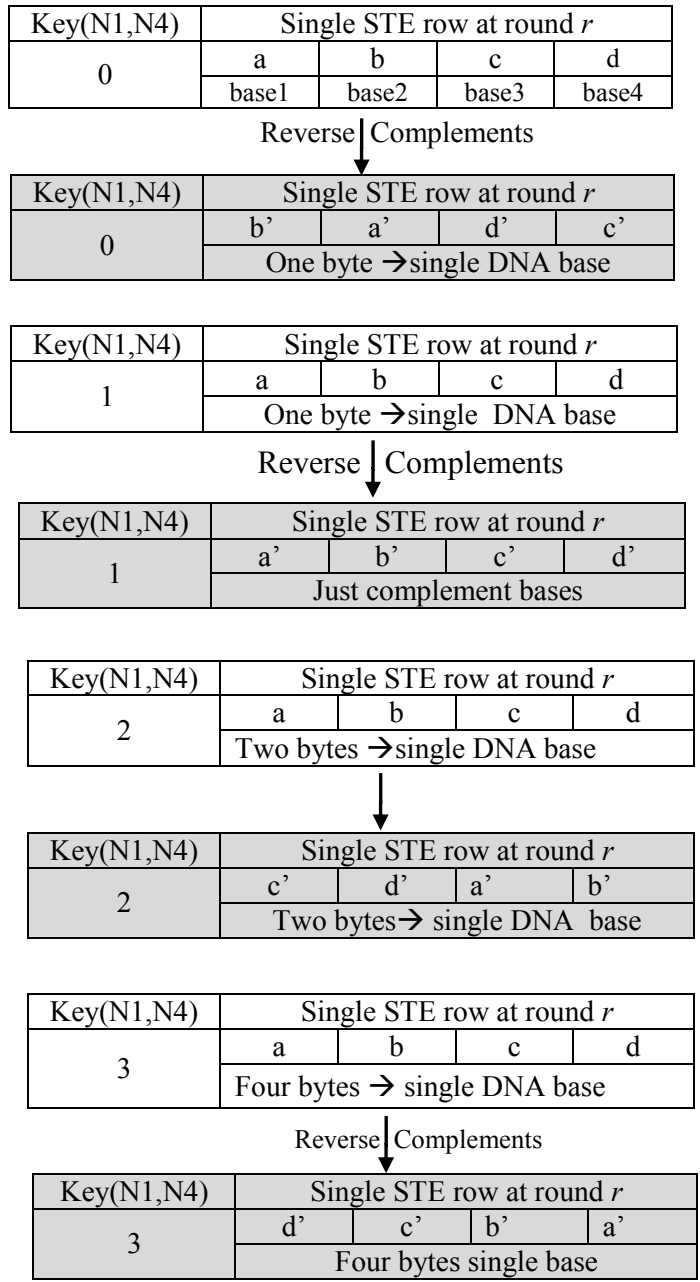

Figure 3. The state row transposition process depending on key

The whole method for the $KdD\_T_r$ at round $r$ using the key cipher $Kr$ according to the process described in Figure 3 is illustrated in Figure 4.

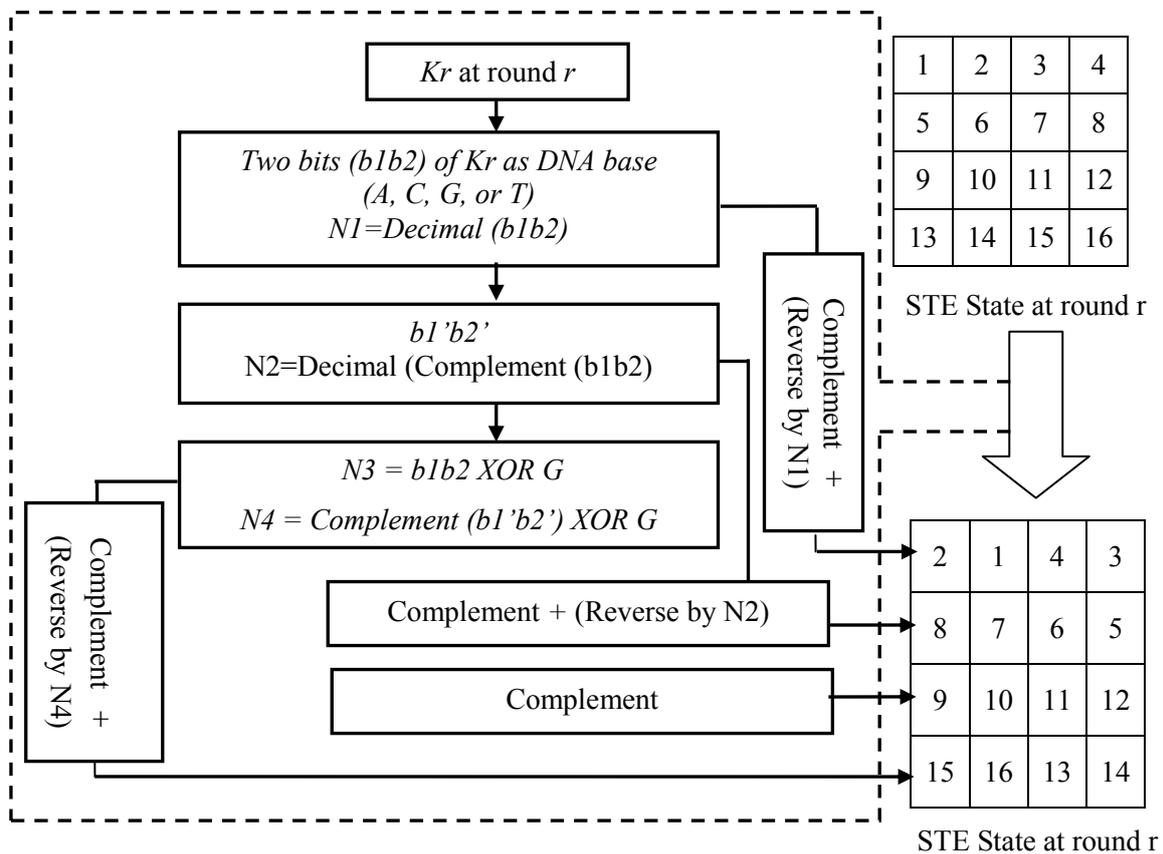

Figure 4. Key-dependent transposition inspired by DNA processes

## 4. TEST AND RESULT

### 4.1 Correlation coefficient

The correlation coefficient assumes values ranging between -1 and +1. According to [10] and [11] the subsequent values are in an agreeable range for explicating the correlation coefficient as declared in Table 2. Note that this paper considers the two variables of plaintext (p) and ciphertext (c)

Table 2. Accepted range values for interpreting the correlation coefficient

| Value | Meaning (state) |
|---|---|
| 0 | Non-Linear relationship |
| +1 | Perfect positive linear relationship |
| -1 | Perfect negative linear relationship : When *p* increases in its values, *c* decreases in its values by an exact linear rule |
| (0,0.3) ,(0,-3) | Weak positive (negative) linear relationship by unstable linear rule. |
| (0.3,0.7) ,(0.3,-7) | Moderate positive (negative) linear relationship. |
| (0.7, 1), (-0.7,-1.) | strong positive (negative) linear relationship |

The following equation describes the use of the correlation coefficients functions:

$$E(c) = \frac{1}{s} \sum_{i=1}^{s} p_i \qquad (1)$$

where: s is the total number of bits, $p_i$, $c_i$ are the series of s measurements for p and c, p is bits value of original bits or plaintext, c is bits value of change bits or ciphertext, E(c) is mathematic expectation of c.

The variance of p can be represented within this equation:

$$D(p) = \frac{1}{s}\sum_{i=1}^{s}[p_i - E(p)]^2 \qquad (2)$$

Finally, the related coefficients $r_{pc}$ can be described in this equation:

$$r_{pc} = \frac{E\{[p - E(c)][c - E(c)]\}}{\sqrt{D(p)}\sqrt{D(c)}} \qquad (3)$$

With a view to test the coefficient of block cipher, the experiment was separated into two different types of testing which are:

Test on permutation transformation only.
Tests on all transformations of whole block cipher which used the proposed permutation transformation.

### 4.1.1 Correlation coefficient on permutation transformation

To evaluate the diffusion property of the block cipher byte permutation, the experiment was examined $KdD\_T_r$ transformation in the AES block cipher. Figure 6 illustrates the laboratory experiment process of correlation coefficient of the $KdD\_T_r$ transformation in the AES block cipher.

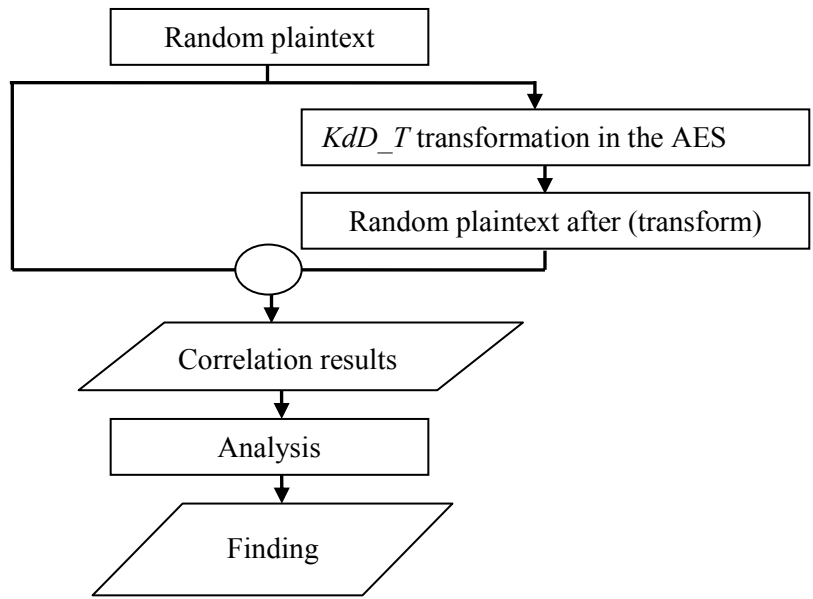

Figure 6. Laboratory experiment process of correlation coefficient $KdD\_T_r$ function.

The analysis on the data from Table 3 is accomplished, in which the correlation value for the plaintext before and after the $KdD\_T_r$ function in the block cipher denoted as newT are recorded.

A scatter chart of the results is presented in Figure 7. It shows that there is no sequences from the $KdD\_T_r$ function in AES block recorded the correlation values between 0.7 and 1.0 which indicates a weak positive (or negative) linear relationship. It also reported that 42 of 128 sequences from the $KdD\_T_r$ function in AES block cipher recorded the correlation values between 0.3 and 0.7, which indicates that (32%) percentage of correlation has a moderate non-linear relationship of the AES block cipher for $KdD\_T_r$ transformation.

Table 3. Correlation test, $r_{xy}$ of byte permutation function for sequence number 1- 128

| #Seq. | newT | #Seq. | newT | #Seq. | newT | #Seq. | newT | #Seq. | newT |
|---|---|---|---|---|---|---|---|---|---|
| 1 | 0.08597 | 29 | 0.18569 | 57 | 0.08363 | 85 | 0.12985 | 113 | 0.18390 |
| 2 | 0.17037 | 30 | 0.40374 | 58 | 0.15349 | 86 | 0.06595 | 114 | 0.03909 |
| 3 | 0.12791 | 31 | 0.23373 | 59 | 0.38738 | 87 | 0.01376 | 115 | 0.16758 |
| 4 | 0.02958 | 32 | 0.19055 | 60 | 0.32680 | 88 | 0.00464 | 116 | 0.50120 |
| 5 | 0.52840 | 33 | 0.10944 | 61 | 0.40326 | 89 | 0.35249 | 117 | 0.22640 |
| 6 | 0.09531 | 34 | 0.37762 | 62 | 0.12639 | 90 | 0.28854 | 118 | 0.03744 |
| 7 | 0.03137 | 35 | 0.05647 | 63 | 0.25217 | 91 | 0.39981 | 119 | 0.05677 |
| 8 | 0.04419 | 36 | 0.13193 | 64 | 0.19812 | 92 | 0.05009 | 120 | 0.34001 |
| 9 | 0.40574 | 37 | 0.31592 | 65 | 0.28045 | 93 | 0.11119 | 121 | 0.18795 |
| 10 | 0.15235 | 38 | 0.58460 | 66 | 0.18487 | 94 | 0.36348 | 122 | 0.30822 |
| 11 | 0.38431 | 39 | 0.47493 | 67 | 0.02447 | 95 | 0.36977 | 123 | 0.38112 |
| 12 | 0.12620 | 40 | 0.29327 | 68 | 0.10077 | 96 | 0.11773 | 124 | 0.10533 |
| 13 | 0.20651 | 41 | 0.06815 | 69 | 0.11906 | 97 | 0.08692 | 125 | 0.28302 |
| 14 | 0.48903 | 42 | 0.27067 | 70 | 0.02194 | 98 | 0.18529 | 126 | 0.36338 |
| 15 | 0.37696 | 43 | 0.18614 | 71 | 0.24791 | 99 | 0.34679 | 127 | 0.13048 |
| 16 | 0.14827 | 44 | 0.19187 | 72 | 0.16198 | 100 | 0.18237 | 128 | 0.00430 |
| 17 | 0.33981 | 45 | 0.45570 | 73 | 0.24469 | 101 | 0.31952 | | |
| 18 | 0.10116 | 46 | 0.02143 | 74 | 0.49951 | 102 | 0.07563 | | |
| 19 | 0.07355 | 47 | 0.13180 | 75 | 0.50552 | 103 | 0.35611 | | |
| 20 | 0.32371 | 48 | 0.18264 | 76 | 0.03400 | 104 | 0.47755 | | |
| 21 | 0.02836 | 49 | 0.15319 | 77 | 0.09519 | 105 | 0.01176 | | |
| 22 | 0.28641 | 50 | 0.39863 | 78 | 0.12623 | 106 | 0.49694 | | |
| 23 | 0.12624 | 51 | 0.07359 | 79 | 0.26368 | 107 | 0.45109 | | |
| 24 | 0.10645 | 52 | 0.16813 | 80 | 0.11925 | 108 | 0.06314 | | |
| 25 | 0.06252 | 53 | 0.21356 | 81 | 0.07441 | 109 | 0.35047 | | |

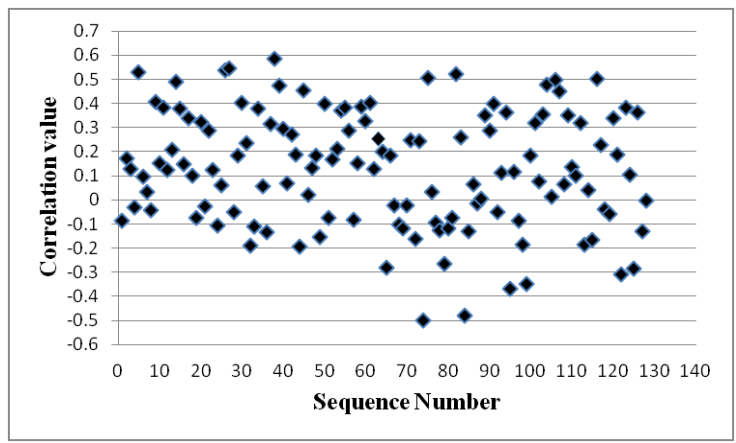

Figure 7. Scatter chart of the correlation test results on $KdD\_T_r$ function only

From the result analysis, it can deduce that this function has a good confusion performance between the plaintext and the ciphertext. This infers that it has a good security.

### 4.1.2 Correlation coefficient on all transformations

The experiment examined all functions of the block cipher. The analyses of the data, in which correlation value for round 1,2 and 3 for each sequence is recorded.

Scatter chart of the results is presented in Figure 8. It illustrates that the majority of correlation values, at different rounds through the whole block cipher are near to 0, which indicates a strong positive (or negative) non-linear relationship. Only 0.007% is near to +1 or -1, in round 1, representing a weak positive (or negative) non-linear relationship. From the results, it can be inferred that the block cipher has an improved confusion performance.

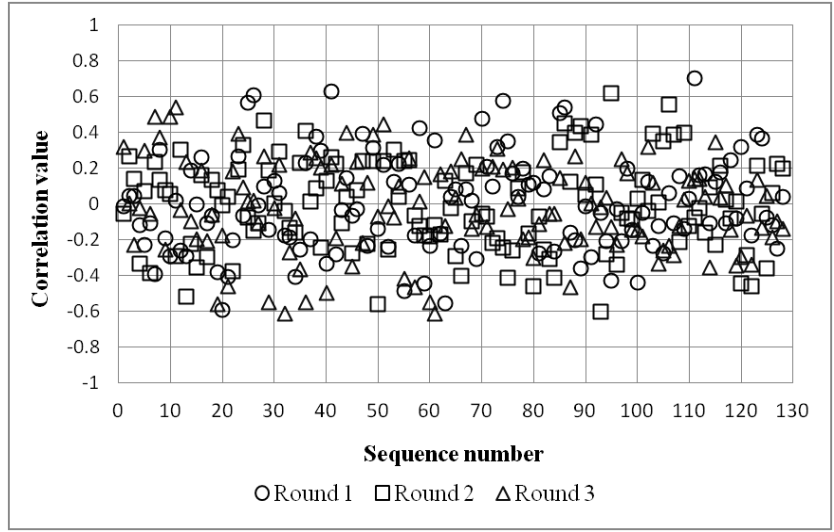

Figure 8.Scatter chart of the correlation test results on whole block cipher

### 4.2 NIST Suite Randomness test

The randomness test is one of the security analysis to measure the confusion and the diffusion properties of the new encryption algorithm , as carried out in [12-14] [15] and [16].

NIST Suite [17] is a statistical test suite for randomness by NIST used to assess the cryptography algorithm. The suite test assesses whether the outputs of the algorithms under certain test condition exhibits properties that would be expected randomly generated outputs.
In order to evaluate the randomness of ciphertext, an experiment that included a set of data as random plaintext and random 16 byte key in the ECB mode was conducted. The 128 sequences of 1,059,584 bits were constructed and examined.

The proportion of sequences that passed a specific statistical test should lie above the proportion value p, defined in Equation (4).
as:

$$p_\alpha = (1-\alpha) - 3\sqrt{\frac{\alpha(1-\alpha)}{n}} = \qquad (4)$$

where $\alpha$ is the significant value, $n$ is the number of tested sequences.
For this experiment, the proportion value is:

$$p_\propto = (1 - 0.01) - 3\sqrt{\frac{0.01(1-0.01)}{128}} = 0.963616 \quad (5)$$

where n= 128, and ∝= 0.01.

The list of statistical tests applied throughout the experiments is illustrated in Table 4.

Table 4. Breakdown of 15 statistical tests applied during experimentation

| Test ID | NIST Statistical Test | Number of p- |
|---|---|---|
| 1 | Frequency | 1 |
| 2 | Frequency-Within-Block | 1 |
| 3 | Runs | 1 |
| 4 | Longest Runs of Ones | 1 |
| 5 | Binary Matrix Rank | 1 |
| 6 | Discrete Fourier | 1 |
| 7 | Non-Over Lapping Template | 1 |
| 8 | Overlapping Template | 1 |
| 9 | Maurer's Universal | 1 |
| 10 | Linear Complexity | 1 |
| 11-12 | Serial | 2 |
| 13 | Approximating Entropy | 1 |
| 14-15 | Cumulative Sums (Cusums) | 2 |
| 16-23 | Random Excursions | 8 |
| 24-41 | Random Excursions Variant | 18 |

The p-value readings for each round constructed is illustrated in Figure 9. This figure demonstrates the randomness test for 15 statistical tests of block cipher that used the proposed

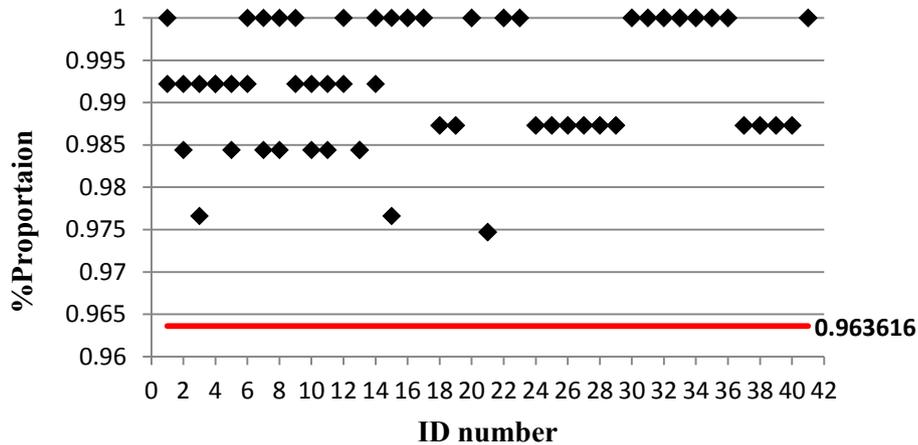

Figure 9. Randomness tests results of whole block cipher in round 2&3

ShiftRows for the rounds 2 and 3, respectively. From this figure, at the end of the second and third rounds, all of the 41 statistical tests fall over 96.3616%, which is evidence that the output from the algorithm is completely random.

## 4.3 CRYPTANALYSIS

The new permutation transformation is dynamic and its changing at each round according to round key-value this feature make the job more hard for the attackers since the analysis of dynamic unit is more difficult than the static one. For the dynamic ShiftRows the attacker have the possibility of $2^n!$ which consider a high number, that increasing the resistance of the algorithm against the attacks.

## 5. CONCLUSION

In this a new dynamic key-dependent permutation transformation was developed with chosen byte from key and existing AES state. The new ShiftRows transformation is not fixed, but changeable at each round according to the round key values. This permutation unit was tested with the correlation coefficient and the 15 statistical randomness tests of NIST Test Suite. Analyzing the results demonstrates that the characteristics of the new ShiftRows are more secure and hard the job for attackers, on which concluded that it is potential to employ it for an encryption. This will increases the stability of AES against linear and differential cryptanalysis. An algorithm for designing this permutation transformation is proposed, as it is based on an inspiration of biology DNA operations.

**Authors**

**Auday H.Al-Wattar** obtained his B.Sc. degree in Computer Science from Mosul University and his M.Sc. Degree from Technology University-Baghdad in 2005 .He presently pursing his Ph.D. From Universiti Putra Malaysia, Malaysia under the guidance of Prof. Dr. Ramaln bin Mahmod at Computer Science and Information Technology Faculty. He works as lecturer at Mosul University (since 2005), at Computer Science and Mathematics Faculty - Computer Science Department. His area of interest includes Computer Security, Programming languages,Data base management. 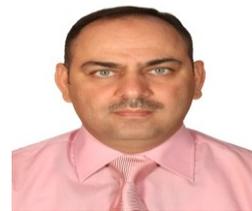

**Ramlan B Mahmod** obtained his B.Sc. in computer Science, from Western Michigan, University, U.S.A. in 1983, and M.Sc. degree in Computer science, from Central Michigan, University, U.S.A. and Ph.D. degree in Artificial Intelligence from, Bradford University, UK in 1994. His previous workings Experience/Position are as:
**System Analyst,** PETRONAS, 1979-1980. **Lecturer,** Mathematic Department, Faculty of Science, UPM, 1985-1994.
**Lecturer,** Department of Multimedia, Faculty of Computer Science & Information Technology, UPM, 1994 – 2002. **Associate Professor,** Department of Multimedia, Faculty of Computer Science & Information Technology, UPM, 2002 – 2010.
**Deputy Dean,** Faculty of Computer Science & Information Technology, UPM, 1998 – Nov 2006. **Dean,** Faculty of Computer Science & Information Technology, UPM, 2010-2013. Professor, Faculty of Computer Science & Information Technology, UPM, 2012 - now. His research interest includes Neural Network, Artificial Intelligence, Computer Security and Image Processing. 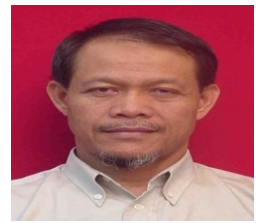

**Nur Izura binti Udzir** obtained her B.Sc. in computer Science, from Universiti Pertanian Malaysia, in 1995, and M.Sc. degree in Computer science, from Universiti Putra Malaysia, in 1998, and Ph.D. degree in Computer Science from, University of York, UK in 2006.
**Associate Professor, Head** Department of Computer Science, Faculty of Computer Science & Information Technology, UPM. 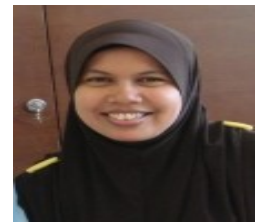

Her research interest includes Computer security, secure operating systems, Access control, Distributed systems, Intrusion detection systems.

**Zuriati Binti Ahmad Zukarnain** obtained her B.Sc. in Physics,, from Universiti Putra Malaysia, in 1997, and M.Sc. degree in Information Technology, from Universiti Putra Malaysia, in 2000, and Ph.D. degree in Quantum Computation and Quantum Information from, University of Bradford, UK, 2006,UK in 2006. 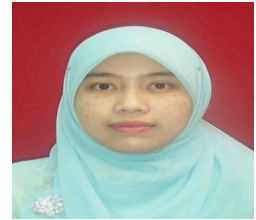
**Associate Professor,** Department Of Communication Technology and Networking, Faculty of Computer Science & Information Technology, UPM. Her research interest includes information, computer and communication technology (ICT), Quantum Information Systems and Distributed Systems, Quantum Computing, Computer Networks and Distributed Computing.